# Variability estimation in a non-linear crack growth simulation model with controlled parameters using Designed Experiments testing

Seungju Yeo[a], Paul Funkenbusch[a], Hesam Askari[a]

[a] Department of Mechanical Engineering, University of Rochester, Rochester, NY 14627, USA

email:
Seungju Yeo (syeo2@ur.rochester.edu) **[corresponding author]**
Paul Funkenbusch (pfunken2@ur.rochester.edu)
Hesam Askari (askari@rochester.edu)

Data availability:
The data that support the findings of this study are available from the corresponding author upon reasonable request.

Funding statement:
The authors declare there is no external funding source to disclose for this study.

Conflict of interest disclosure:
The authors declare there is no conflict of interest to disclose for this study.

**Abstract**

Variability in multiple independent input parameters makes it difficult to estimate the resultant variability in the system's overall response. The Propagation of Errors (PE) and Monte-Carlo (MC) techniques are two major methods to predict the variability of a system. However, in the former method, the formalism can lead to an inaccurate estimate for systems that have parameters varying over a wide range. For the latter, the results give a direct estimate of the variance of the response, but for complex systems with many parameters, the number of trials necessary to yield an accurate estimate can be very large to the point the technique becomes impractical. In this study, the effectiveness of the Tolerance Design (TD) method to estimate variability in complex systems is studied. We use a linear elastic 3-point bending beam model and a nonlinear extended finite elements crack growth model to test and compare the PE and MC methods with the TD method. Results from an MC estimate, using 10,000 trials, serve as a reference to validate the result in both cases. We find that the PE method works suboptimal for a coefficient of variance above 5% in the input variables. In addition, we find that the TD method works very well with moderately sized trials of designed experiment for both models. Our results demonstrate how the variability estimation methods perform in the deterministic domain of numerical simulations and can assist in designing physical tests by providing a guideline performance measure.

**Introduction**

Many engineered designs aim for a controlled level of variability of the system response to achieve the intended purpose of the design [1], though complex engineering systems often have many sources of variability that can significantly affect the outcome of the system. Examples of sources of variability in manufacturing include variations in raw material's properties, details of the manufacturing process and assembly procedures, changes in operating environment, differences in use conditions, and uncertainty in measurements of performance. Variability in multiple input parameters can make it difficult to predict the resultant variability in the overall response of the system [2]. Various statistical techniques, with various levels of sophistication, have been applied to estimate system variability. Examples of these techniques include summation of worst case variations [1], [3], root sum square (RSS) to add up the contributions from precision and bias errors [4], quasi-likelihood estimation [5], Propagation of Error (PE) method [6], Monte-Carlo (MC) method, and Tolerance Design (TD) method. These methods provide a mathematical approach to consider variability of individual parameters and extrapolate how that will affect the variability of the system response. In this study we will compare results from three of these methods, Propagation of Error, Monte Carlo, and Tolerance Design.

The first technique of interest is the PE method. This technique involves determining the partial derivative of the response with respect to each parameter, evaluated at the center point of the parameter space. This is then used to determine the ratio between the variance of the parameter and its contribution to the variance of the response. Adding the contributions from all the parameters provides an estimate of the total variance of the response. As an example, Mount and Louis utilized PE to estimate the error bound in measurements of the lateral movement of river channels from aerial photographs [7]. This technique is straightforward and generally yields a good estimate if the variabilities in the parameters are small [8]. However, because the derivatives are evaluated at the center point, PE method may give a poor estimate if the parameters vary over a wide range for instance in the measurement of elasticity of a steel bar where fracture and damage are present.

The MC technique involves randomly sampling the parameter space based on the frequency distribution of the space [9]. Multiple trials are run with the randomly selected values of each parameter for each trial. The results of these trials give a direct estimate of the variance of the response. This technique is expected to produce excellent "gold standard" estimates of the variance provided enough trials are included. For example, Beck and da Silva used MC method to verify the theory induced from mathematical model of uncertainty propagation of Timoshenko and Euler beam theory [10]. Unfortunately, MC techniques are difficult to use for physical systems because of the considerable number of random trials required. The method may be impractical even for numerical models if the calculation time per trial is too large [9]. Hence, it raises the interest in using a small number of structured trials in the form of a designed experiment.

One simple but effective empirical method for variance estimation is TD approach popularized by Taguchi [1], [9], [11]. This method involves conducting a small number of trials that sample the parameter space in a controlled pattern. This method uses common design of experiments tools (factorial arrays [12], ANOVA [13]), without the need for additional multi-stage testing or modeling to restrain the system response [14]. It can be used on computer models as well as physical systems and allows the effect of changes in input parameter uncertainty (e.g., tightening tolerances, reducing measurement uncertainty) to be predicted. An additional advantage, for physical systems, is that it allows experiments to be conducted with a limited number of distinct settings (levels), which can simplify experimentation. For example, a TD experiment for a throttle handle assembly with seven dimensional parameters which translates to a $2^7$ dimensional space was done by using only 16 properly designed prototypes by Bisgaard [15]. However, although TD is widely used [7], its ability to accurately estimate variance is necessarily dependent on the adequacy of the parameters that are assumed to be significant to the system response, or the underlying correlation among different parameters of the system being studied. Therefore, the performance of this method is yet to be fully understood in the presence of complex mechanical behavior such as failure and crack propagation.

In this study, the effectiveness of using a designed experiment to estimate variability for a non-linear crack propagation problem is examined by comparing the variance estimated from TD method against the variance found using MC simulations. Two mathematical models are used for the comparisons here. The first is a simple beam bending model for which an analytic solution is available. This is done so that the PE methods can be applied. The calculation time for this model is small, which also allows multiple, large MC simulations to be run so that the effects of various levels of input variability can be examined. The simplicity of this model makes it easier to illustrate the application of the different techniques in the absence of structural nonlinearities. The second model is a non-linear extended finite element method (XFEM) model of crack opening in an aluminum plate loaded in tension. This model is much more complex and requires a much longer time to execute, exemplifying how using TD method saves time. In this study, the input parameters of the first model are varied by 1%, 2%, 5%, 10%, and 20% and the results are compared using the PE, MC, and TD methods. Then the input parameters for the second model are varied by 5% as a representative variability limit and compared using the MC and TD approach. The findings from the two models are discussed in detail in sections 3, 4, and 5.

**Materials and Methods**
**Analytical method formalism**
*Propagation of Error*
The PE method estimates the variance in the response by adding a contribution from each of the input parameters. Assuming each parameter is uncorrelated to all others, its contribution to the overall system's response is assumed to be proportional to the variance in the value of the parameter itself. The proportionality constant is estimated using the partial derivative of the function with respect to corresponding variable, evaluated with all parameters at their average values, i.e., at the center point. The estimated variation in the system output can be expressed as follows:

$$S_v = \sqrt{\sum \sigma_i^2 \left(\frac{\partial v}{\partial x_i}\right)^2} \qquad (1)$$

Where $S_v$ is the estimated standard deviation in the response, $\sigma_i$ is the true standard deviation of each parameter and $v$ is the system response as a function of different parameter $x_i$, the corresponding parameter. The analytic equation for the response is required to apply this method to obtain the partial differentiation of the system output with respect to each parameter.

*Monte-Carlo*
The estimate for the standard deviation of the response is:

$$S_v = \sqrt{\frac{\sum_{i=1}^{N}(v_i-\bar{v})^2}{N-1}} \qquad (2)$$

Where N is the number of trials, $v_i$ is the system response for each trial, and $\bar{v}$ is the average value of the response from all trials. [16]

*Tolerance Design*
In the TD approach, each parameter is limited to a small number of distinct values or levels. Like the PE calculations expressed in Equation 1, it assumes that the variance of the response is proportional to the variance of each parameter in the design. Hence, the small number of values of each parameter can be selected so that they can also represent the variance of the corresponding parameter in the original distribution, if the parameter space is continuous and shows any variance. Two values are often enough to represent a parameter space that is assumed symmetric on both sides of the central peak [17]. Each parameter is represented by two levels, a standard deviation above and below the average value, each value occurring 50% of the time. Calculations are then performed for a small number of trials, with the value of each parameter altered between the two levels in a controlled pattern. In this study, five different 2-level designs were tested; 64 (factorial, resolution V), 32 (fractional factorial, resolution IV), 16 (fractional factorial, resolution III), 12 (Placket-Burman design, resolution III), and 8 (fractional factorial, resolution III) [17].

**Application to the three-point beam bending analytical model**
*Model setting*

To illustrate the application of the PE, MC, and TD methods, it is convenient to begin with a system with a simple model that is composed of known parameters and also can be solved analytically as pointed out in section 2.1.1., for direct application of the methods. The model used here is the 3-point bending of a linear elastic Timoshenko beam without any failure criteria, fixed on one end and simply supported on the other end as illustrated in Figure 1.

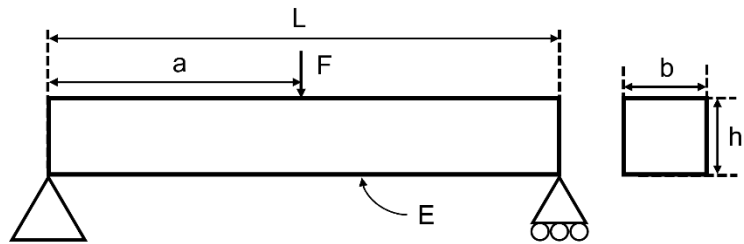

Figure 1. The model setting illustrates a 3-point bending setup of a linear elastic Timoshenko beam, fixed on one end, and simply supported on the other. The beam has a square cross-section with significant dimensions and properties shown.

The maximum vertical displacement, $v$, is chosen to be the system response. The vertical displacement is a function of six parameters according to the analytical solution provided in Equation 3 [18]: the force applied to the beam (F), the distance between the left support and the load point (a), the length of the bar (L), the width of the cross-section (b), the height of the cross-section (h), and Young's Modulus of the beam (E).

$$v = \frac{2Fa(L-a)}{bh^3 LE}(L^2 - a^2 - (L-a)^2) \qquad (3)$$

As a physical example of deflection of such a beam under loading, we can consider a wooden balance bar with square cross section supporting the weight of a person standing at the middle. If we are to conduct different experiments with several beams and different volunteers to acquire the system response of the vertical displacement, the outcome will be different each time, because there will be a variance in each parameter – the dimension of the beam, the weight of the person, or the location of the person on the beam. To observe how the PE, MC, and TD methods estimate the variance of the system response, each parameter is assumed to be continuous, with a nominal value and a variance. When all parameters are set to their nominal values as reported in Table 1, the analytical value of the vertical displacement is 0.196 mm. The variance of each parameter is discussed in the following details of the variation estimation methods.

In an actual application, each parameter would have a different variability associated with it.

For instance, the variability of the weight of the person might be larger compared to the variability of the dimensions of the beam. However, to illustrate how the degree of the uncertainty of the parameters influences the result with PE, MC or TD method, identical coefficients of variance, defined as the ratio of the standard deviation to the nominal value, are used for all parameters in each calculation for the three methods. Five different coefficients of variance are tested: 1%, 2%, 5%, 10% and 20%.

| Parameters | Nominal Value |
|---|---|
| F | 785 N |
| a | 50 cm |
| L | 1.0 m |
| b | 10 cm |
| h | 10 cm |
| E | 10 GPa |

Table 1. Nominal values of parameters based on a physical example of deflection of a wooden linear elastic Timoshenko balance beam under 3-point bending with square cross section supporting the weight of a person standing at the center. Each parameter symbolizes the force applied to the beam due to the weight of the person (F), the distance between the left support and the load point (a), the length of the bar (L), the width of the cross-section (b), the height of the cross-section (h), and Young's Modulus of the beam (E).

*Propagation of Error Analysis*

Consider the model being regenerated multiple times with a statistical variation in each parameter. Then the vertical displacement will also show a distribution of values because it is a function of the parameters. The variance of the vertical displacement is estimated using the PE equation as follows:

$$S_v^2 = \sigma_F^2 \left(\frac{\partial v}{\partial F}\right)^2 + \sigma_a^2 \left(\frac{\partial v}{\partial a}\right)^2 + \sigma_L^2 \left(\frac{\partial v}{\partial L}\right)^2 + \sigma_b^2 \left(\frac{\partial v}{\partial b}\right)^2 + \sigma_h^2 \left(\frac{\partial v}{\partial h}\right)^2 + \sigma_E^2 \left(\frac{\partial v}{\partial E}\right)^2 \qquad (4)$$

Where $\sigma$ is the standard deviation and $\sigma^2$ is the variance of each parameter. The partial derivatives are calculated from the governing equation with respect to each parameter. The weight of each parameter's contribution to the final value is evaluated at the average value as shown in Table 2. As an example, the estimated standard deviation of the vertical displacement with the coefficient of variance at 2% for each parameter is 0.0180 mm, which is about 9.2% of the theoretical result of 0.196mm. Results obtained for PE using Equation 3 and 4 are shown in section 3.1.

| Parameter | Partial Derivative | Numerical value |
|---|---|---|
| F | $\dfrac{4a^2(L-a)^2}{bh^3LE}$ | $2.5 \times 10^{-7}\ m/N$ |
| a | $\dfrac{8a(2a^2 - 3aL + L^2)F}{bh^3LE}$ | $0\ (unitless)$ |
| L | $\dfrac{4a^2(L^2 - a^2)F}{bh^3L^2E}$ | $5.8875 \times 10^{-4}\ (unitless)$ |
| b | $-\dfrac{4a^2(L-a)^2 F}{b^2h^3LE}$ | $-1.9625 \times 10^{-3}\ (unitless)$ |
| h | $-\dfrac{12a^2(L-a)^2 F}{bh^4LE}$ | $-5.8875 \times 10^{-3}\ (unitless)$ |

| | | | |
|---|---|---|---|
| E | $-\dfrac{4a^2(L-a)^2 F}{bh^3 L E^2}$ | | $-1.9625 \times 10^{-14}\ m^3/N$ |

Table 2. The partial derivative of vertical displacement with respect to each parameter and their numerical value at the average

*Monte-Carlo Analysis*

The MC method is executed with an iteration of five coefficients of variance using MATLAB R2021b (Copyright © 2021 MathWorks Inc.). For each iteration of the loop, the nominal value of each parameter multiplied by the coefficient of variance is set as the standard deviation of the corresponding parameter space. Then a constant for each parameter is chosen within the range of one standard deviation below and above the nominal value. Then the parameters are substituted into Equation 3 to yield the calculated vertical displacement of the three-point bending beam for that trial. This process is repeated to obtain various sizes of sample distribution with an average and a standard deviation. The results are shown in section 3.1.

*Tolerance Design Analysis*

As an example, Table 3 shows each parameter when coefficient of variance is 2% with the two representative values, one standard deviation magnitude above and below the average, labeled as level +1 or level -1, respectively. Table 4 shows values used in each trial for the fractional factorial design with array size of 8 trials. The variance of response for each design is acquired from the summarization using the Fit Model feature of JMP® Pro ver. 16.0.0 (Copyright © 2021 SAS Institute Inc.) and the results are shown in section 3.1.

| Parameters | Level -1 | Nominal Value | Level +1 |
|---|---|---|---|
| F | 769.3 N | 785 N | 800.7 N |
| a | 49 cm | 50 cm | 51 cm |
| L | 0.98 m | 1 m | 1.02 m |
| b | 9.8 cm | 10 cm | 10.2 cm |
| h | 9.8 cm | 10 cm | 10.2 cm |
| E | 9.8 GPa | 10 GPa | 10.2 GPa |

Table 3. Nominal values of the parameters and the values corresponding to the upper (+1) and lower (-1) levels when coefficient of variance is 2%. The same assumptions are made for the nominal values as in Table 1.

| Trial | F (N) | a (m) | L (m) | b (m) | h (m) | E (GPa) |
|---|---|---|---|---|---|---|
| 1 | 800.7 | 0.49 | 0.98 | 0.102 | 0.102 | 9.8 |
| 2 | 800.7 | 0.51 | 1.02 | 0.102 | 0.102 | 10.2 |
| 3 | 769.3 | 0.49 | 0.98 | 0.102 | 0.098 | 10.2 |
| 4 | 800.7 | 0.51 | 0.98 | 0.098 | 0.098 | 10.2 |
| 5 | 769.3 | 0.51 | 1.02 | 0.102 | 0.098 | 9.8 |
| 6 | 769.3 | 0.51 | 0.98 | 0.098 | 0.102 | 9.8 |
| 7 | 769.3 | 0.49 | 1.02 | 0.098 | 0.102 | 10.2 |
| 8 | 800.7 | 0.49 | 1.02 | 0.098 | 0.098 | 9.8 |

Table 4. Parameter values used in each trial for the fractional factorial design with array size 8 at the coefficient of variance of 2%. Each trial shows a unique combination of the upper or lower level of each parameter, thus yielding a unique response.

**Application to the non-linear crack growth simulation model**
*Model setting*

The same techniques are applied to the non-linear simulation model of crack growth on an aluminum plate subject to tensile displacement, as illustrated in Figure 2. The measured output is the crack opening displacement in the vertical direction across the crack at the final displacement on the top of the plate.

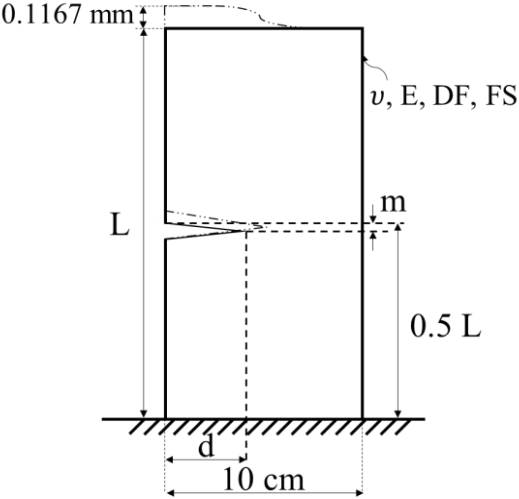

Figure 2. The model setting illustrates an aluminum plate with a notch on the longer edge under tensile displacement. The plate is fixed on one end, and a constant displacement of 0.1167 mm is applied at the half of the top surface, which is the side closer to the notch. The significant dimensions and properties are shown: the length of the aluminum plate (L), initial depth of the crack (d), offset location of the crack (m), Poisson's ratio ($v$), Young's Modulus (E), Distance at Failure (DF), and Failure Strength (FS). The deformed shape of the plate is presented by lines with long dash and dots.

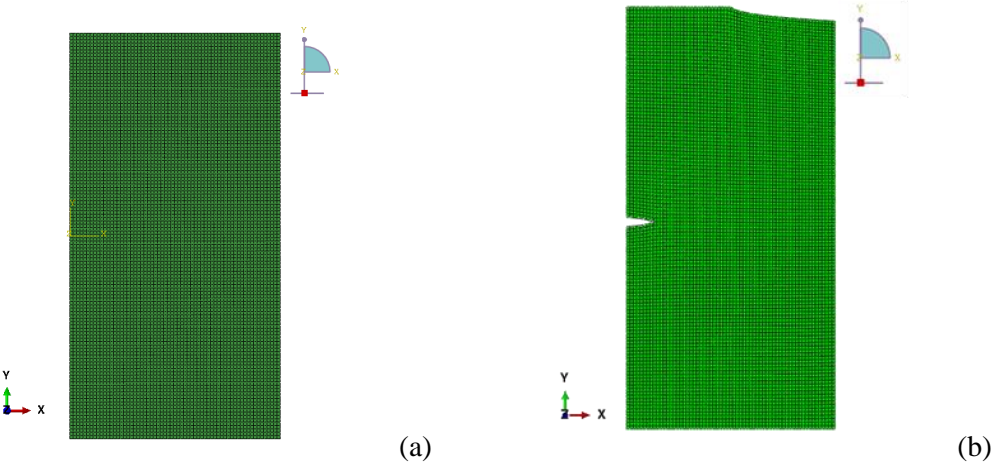

(a)  (b)

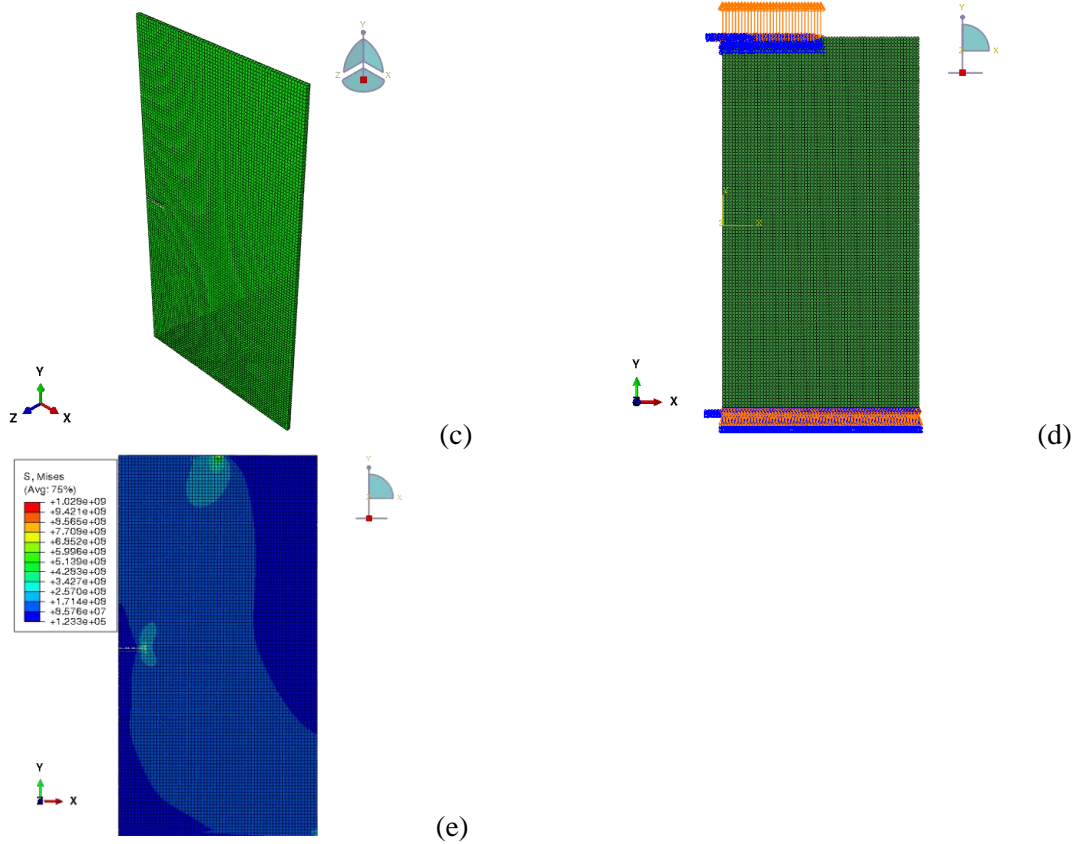

Figure 3. (a) X-Y plane view of Finite Element Model of aluminum plate (b) Deformed model with a scale factor of 100 in y direction. (c) Perspective view of the plate. (d) Model view of the plate with the boundary conditions. The fixed constraint is created at the initial step at the bottom edge of the plate. Displacement of 0.1167 mm to positive y direction is applied at the top left half of the edge, created at the secondary step. (e) Von-Mises stress contour plot of the plate at the final step.

An XFEM non-linear crack simulation model is generated with Abaqus/Explicit (Copyright © 2020 Dassault Systèmes). The crack domain that contains a crack is a thin aluminum plate 10 cm wide, 20 cm long and 0.2 cm thick. The initial crack is located slightly off the center of the longer edge of the plate. The bottom face is fixed, and half of the top face is pulled off to open the crack. A constant displacement of 0.1167 mm is applied on the top surface. The model configuration is shown in Figure 3. The opened length of the crack along the long edge is a function of many unknown parameters. Among those, seven parameters are assumed to have a significant level of uncertainty in their value that affects the final crack opening displacement: the length of the aluminum plate (L), initial depth of the crack (d), offset location of the crack (m), Poisson's ratio ($\upsilon$), Young's Modulus (E), Distance at Failure (DF), and Failure Strength (FS).

| Parameters | Level -1 | Nominal Value | Level +1 |
|---|---|---|---|
| L | 19 cm | 20 cm | 21 cm |
| d | 0.95 cm | 1 cm | 1.05 cm |
| m | 0.95 mm | 1 mm | 1.05 mm |
| $\upsilon$ | 0.285 | 0.3 | 0.315 |
| E | 190 GPa | 200 GPa | 210 GPa |
| DF | 0.095 mm | 0.1 mm | 0.105 mm |
| FS | 380 MPa | 400 MPa | 420 MPa |

Table 5. Nominal values of the parameters and the values corresponding to the upper (+1) and lower (-1) levels with 5% of coefficient of variance. The parameters represent the length of the aluminum plate (L), initial depth of the crack (d), offset location of the crack (m), Poisson's ratio ($v$), Young's Modulus (E), Distance at Failure (DF), and Failure Strength (FS). The nominal values are chosen based on a physical model of a thin aluminum plate containing a crack subjected to tensile displacement. The initial crack is located slightly off the center of the longer edge of the plate.

The thickness and width of the plate, displacement applied on the top face, and density of the material are kept constant. The thickness of the plate is set to a constant to ensure the plate is one mesh element thick so that the crack opening displacement is symmetric along the thickness. The mesh element type used in the study is C3D8 which is a linear cuboid with a node at each corner, and the mesh size is 0.16 cm over the entire model. Both the mesh element type and the mesh size are chosen considering the need to minimize computation time and convergence as detailed in the supplementary materials. The applied displacement is chosen such that the crack propagates without breaking the material into two pieces. The measured crack opening displacement is the largest difference between the displacement of six surrounding nodes across the notch. Six nodes are chosen to ensure that there are enough nodes to capture the actual crack displacement while maintaining the robustness of the experiment using many cases. Figure 4 shows an example of the chosen nodes around the notch.

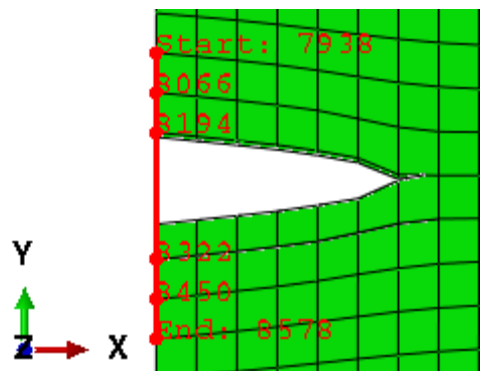

Figure 4. Example of node path of six nodes around the crack with an exaggerated deformation in y direction with scale factor of 100. The numbers in red indicate the node number of the corresponding node. Six nodes are chosen to ensure that there are enough number of nodes to capture the actual crack displacement while maintaining the robustness of the experiment using many cases.

*Monte-Carlo Analysis*

For the MC model, each of the seven parameters is a random constant that follows a Gaussian distribution with a standard deviation of 5% of the nominal value, so that each model contains a unique set of parameters. The 10,000 crack opening displacements from the MC model are collected to plot a distribution curve and acquire the standard deviation.

*Tolerance Design Analysis*

As in the analytic model, the values of seven parameters are set one standard deviation away from the average, so that the variance of the parameter in the model is equal to that in the full Gaussian distribution. The size of orthogonal arrays used are 128 (factorial, full resolution), 64 (factorial, resolution IV), 32 (factorial, resolution III), 16 (factorial, resolution III), 12 (Placket-Burman design), and 8 (factorial, resolution III). [17] The arrays are generated using JMP® Pro ver. 16.0.0 (Copyright © 2021 SAS Institute Inc.).

## Results

*Analytical model*

Monte Carlo results are obtained for sample sizes of 10, 20, 50, 100, 500, 1,000, and 10,000, with multiple calculations done for each sample size N. The results obtained with N = 10,000 is used as the true value for comparison with the other results. As shown in Figure 5, the standard deviation is scattered but converges as the sample size increases toward 10,000.

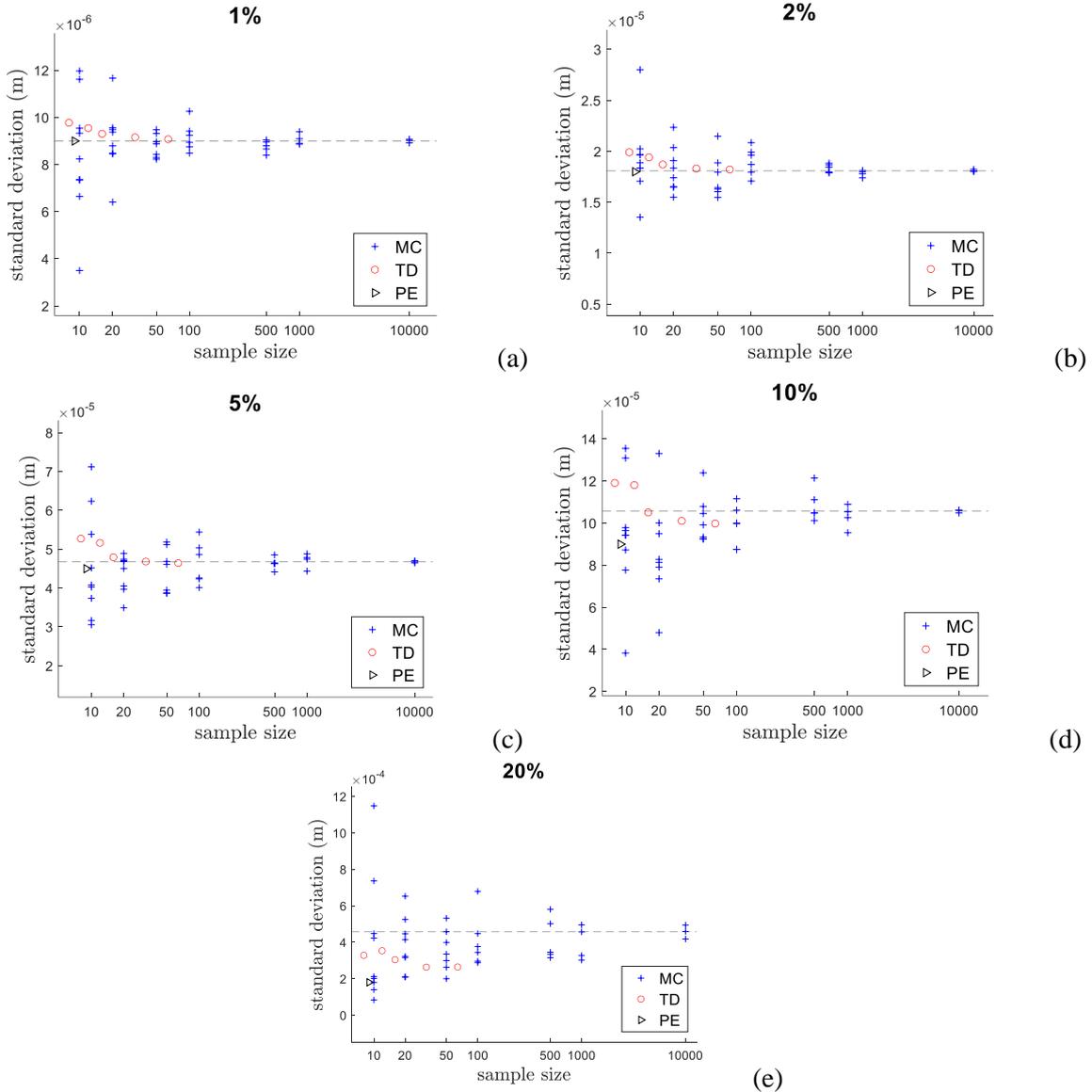

Figure 5. Distribution of standard deviation estimated from Monte-Carlo (blue crosses), Propagation of Error (black triangle), and Tolerance Design (red circles) methods for different sample sizes (horizontal axis) with coefficient of variation of (a) 1%, (b) 2%, (c) 5%, (d) 10%, and (e) 20% for the input parameters. The dashed line indicates the average of the standard deviations obtained from Monte-Carlo calculations with a sample size of 10,000. Each plot shows the results for a different overall level of variability.

Using the N = 10,000 MC values as the standard for comparison, both the PE and TD methods produce better estimates of the variability than random MC calculations with much larger sample sizes as shown in Figure 5. An exception occurs at the highest level of variability examined where none of the techniques does well. The relative error of each method is shown in Table 6.

|        | 1%    | 2%    | 5%     | 10%    | 20%    |
|--------|-------|-------|--------|--------|--------|
| PE     | 0.06% | 0.58% | 3.75%  | 14.83% | 60.62% |
| TD (8) | 8.55% | 9.99% | 12.80% | 12.69% | 28.18% |
| TD (12)| 6.02% | 7.23% | 10.44% | 11.75% | 22.71% |
| TD (16)| 3.35% | 3.36% | 2.52%  | 0.57%  | 33.44% |
| TD (32)| 1.68% | 1.15% | 0.17%  | 4.35%  | 42.42% |
| TD (64)| 0.79% | 0.60% | 0.69%  | 5.58%  | 42.20% |

Table 6. Relative error of the Propagation of Error method (PE) and the Tolerance Design method (TD) with different array sizes at each coefficient of variance with respect to Monte-Carlo method. The array size of each TD method is presented in parenthesis.

In Figure 6, standard deviation results from the PE and TD methods are normalized against the N = 10,000 value. PE method does as well or better than the TD method for small values of the coefficient of variance. In the intermediate range of coefficient of variance from 5% to 10%, the TD method begins to outperform the PE method. The larger arrays (N = 16, 32, 64) of TD also tend to outperform smaller arrays (N = 8, 12), although this breaks down when the coefficient of variance is 20%, where none of the methods does well.

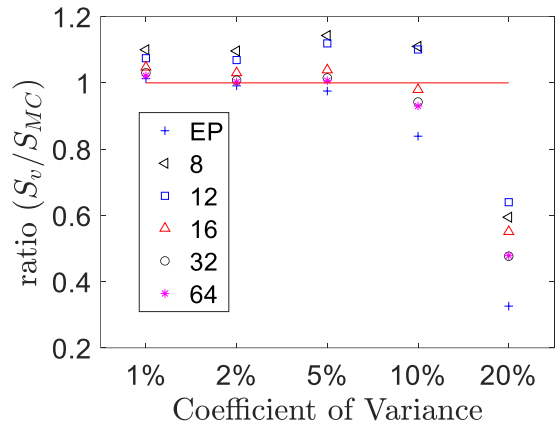

Figure 6. Estimated standard deviation versus coefficient of variance using Propagation of Error and different Tolerance Design arrays. The standard deviations are normalized against the value obtained from Monte-Carlo calculations with N = 10,000. The red line indicates the ratio of the standard deviation of Monte-Carlo model to itself as a reference.

*Simulation model*

The standard deviation of the TD method is shown in Table 7. The distribution of 10,000 cases in the MC model yielded an average of 4.0934e-05 m, and the standard deviation of 4.0811e-06 m. The final crack opening distance distribution of MC model is shown in Figure 7. The distribution has two main peaks. The gross sum is symmetric below and above the mode. Figure 8 shows the ratio of standard deviation of TD to those of N = 10,000 MC method. It is seen that the estimate of the variability is better with larger arrays. The relative error of standard deviation varies from 16.09% for an array of 12 to -1.53% for an array size of 32.

| Array Size | Standard Deviation ($\times 10^{-6}\ m$) |
|---|---|
| 8  | 4.609047 |
| 12 | 4.732711 |
| 16 | 4.430932 |

| 32 | 4.014394 |
| 64 | 3.981951 |
| 128 | 3.966286 |

Table 7. Standard deviation value of different array design of the TD method

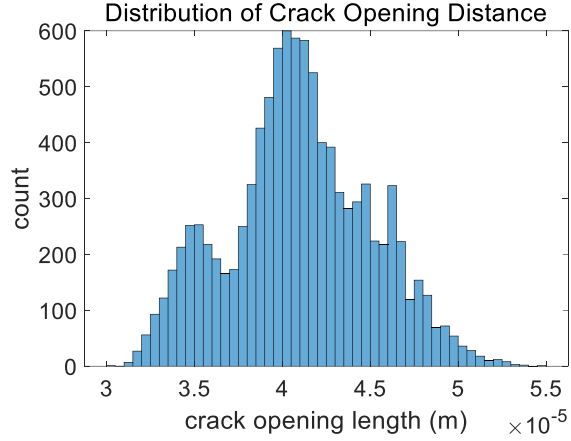

Figure 7. Histogram of crack opening distance of MC model. This bimodal distribution is slightly skewed to the right. The most frequent bin ranges from $4.00 \times 10^{-5}$m to $4.05 \times 10^{-5}$m with the count of 600 out of 10,000 cases. The median is $4.20 \times 10^{-5}$m and the mean is $4.0934 \times 10^{-5}$m. The standard deviation of this distribution is $4.0811 \times 10^{-6}$m.

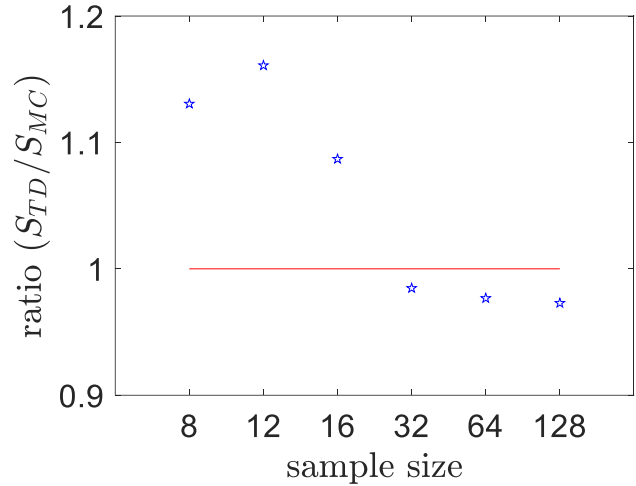

Figure 8. Relative standard deviation of Tolerance Design method at each array design to Monte-Carlo model (blue stars). The red line indicates the ratio of the standard deviation of Monte-Carlo model to itself as a reference.

**Discussion**

In the linear 3-point bending beam model, the output is the vertical displacement of the beam at the load point, determined by six parameters with a constant coefficient of variance at each trial. The beam model is based on a physical model of a wooden linear elastic Timoshenko balance beam with square cross section fixed at one end and simply supported at the other. Without any failure or dynamic response, the weight of a person standing at the center is the only force inducing the displacement of the beam. The six parameters are the force applied to the beam due to the weight of the person, the distance between the left support and the load point, the length of the bar, the width of the cross-section, the height of the cross-section, and Young's Modulus of the beam. With the constant coefficient of variance over different parameters, the standard deviation of the output is computed with the TD, PE, and MC methods. The MC method serves as a reference because it directly estimates the standard deviation without any further manipulation of the raw data.

The results of the TD and PE methods are compared to the results of the MC method to evaluate their effectiveness in the estimation of the standard deviation. The TD and PE methods show good estimation of the variability in the analytic model of a 3-point bending beam, compared to the MC method under different levels of uncertainty, for the lower range of variability tested in this study – 1%, 2% and 5%. The relative error of the estimation of the PE method is 0.06% with 1% of the coefficient of variance, and 0.58% with 2% of the coefficient of variance. The results of the TD method have relative error of 0.79% and 0.60% respectively. In addition, the TD method presents better estimation for variability compared to PE in an intermediate range where the coefficient of variance is 5% and 10%. As the coefficient of variance of the parameters increases, the effectiveness of the PE method decreases faster than the TD method does, because the derivatives of the PE method are evaluated at the center point. The TD method with larger array designs yields a better estimation for small variance, but at a certain point further increases in array size do not improve the performance. Any array design yields a better estimation than the smallest array design for any coefficient of variance.

In the non-linear crack growth model, seven parameters are assumed to be affecting the system output of the crack opening displacement on an aluminum plate under tensile stress. The parameters are the length of the aluminum plate, initial depth of the crack, offset location of the initial notch where crack starts, Poisson's ratio, Young's Modulus, Distance at Failure, and Failure Strength. The coefficient of variance is 5% over those seven parameters. The notch is placed along the longer edge of the plate. The tensile displacement of 0.1167 mm is applied to the half of the top surface closer to the notch, and the bottom surface remains fixed. The amount of displacement is decided to avoid the plate breaking into two pieces or the crack not opening at all. The study uses C3D8 as mesh element with the global size of 0.16 cm considering the convergence and computational time. The MC method is again used as the reference to evaluate the performance of the TD method. 10,000 cases of MC model are generated to calculate the standard deviation.

The TD method shows about 13% relative error at array size 8, 16% at size 12, 8.6% at size 16, 1.6% at size 32, 2.4 % at size 64, and 2.8% at size 128. The relative error of the TD method of variability estimation is better with larger array sizes than the smaller array sizes. This is similar to the trend observed in the linear model at 5% variance. At its best performance, the method yields a result with about 1.5% relative error that could be obtained more than 300 times faster compared to MC. This reduction of time becomes more significant as the complexity of the model intensifies and the importance of efficiency from cost to time savings of the experiment rises.

The output of the model is assumed to be affected by seven different parameters, but it is possible that there are parameters kept constant in this study that still play a significant role in determining the opening length of the crack such as the thickness of the plate. Unlike the physical system, the simulation model disregards the common variation sources such as measurement uncertainty. The measurement in the model is always consistent because it is made at the final displacement numerically. It is still left unexplored how effectively the methods would estimate the standard deviation of the result if each parameter had different coefficient of variances at each trial. A common trait of the TD method in both analytical beam model and non-linear crack model is that it tends to show the better performance with larger array sizes than relatively smaller array sizes. However, the performance does not improve as the array size increases at a certain point.

**Conclusion**

In this study, the effectiveness of estimating the variance using the Tolerance Design (TD) method, as well as the Propagation of Errors (PE) technique, is validated against a large sample Monte-Carlo (MC) simulation for both a simple beam bending model and a more complicated crack simulation model. The modeling environment allows for a precise and confined set of variable parameters and eliminates the concern of unaccounted parameters affecting the outcomes. Thus, provides us with a set of data that is deterministic within the bounds of the varied parameter space. Taking advantage of its trivial calculation time and simple analytical solution, the application of the PE, MC and TD models are compared for the beam bending model. We find that the PE method predicts the variance well when the coefficient of variance is small, but the TD method has a better estimation for variability in an intermediate range. Larger array sizes yield better estimation for small or intermediate variability. The TD method with the larger array size generally has better estimates than the smaller array size. We compared the TD and MC methods in the crack simulation model which takes more computational resources without an analytical solution than the previous model. We find that the relative error is in the range of 8 % to 16 % for smaller array designs of the TD method, but the error decreases to less than 3% for larger array designs with resolution higher than Ⅲ. The relative error of the TD method is better with higher resolution than the lower resolution. The future work includes efficiency study of the TD method with lower variability of the parameter space. The same study can be repeated with different variability of different parameters at each trial to further investigate the variance estimation ability of the TD method in simulation experiments.

**Conflict of Interest**
The authors declare there is no potential competing interest.

**Acknowledgement**
The authors would like to thank the Mechanical Engineering department at the University of Rochester for the support. The authors also would like to thank Peter Miklavčič and Brendan Mort for their support and feedback.